\documentclass[11pt,aps, showkeys]{revtex4}
\usepackage{amssymb}
\usepackage{amsfonts}
\usepackage{amsmath}   
\usepackage{graphicx}
\usepackage{color}

\newcommand{\elec}{{\cal E}}
\newcommand{\refeq}[1]{(\ref{#1})}
\newcommand{\heavi}{{\Theta}}
\begin{document}

\title{Temperature gradients in equilibrium: small microcanonical systems in an external field}
\author{Alberto Salazar}
\email{albertdenou@gmail.com}
\affiliation{Instituto de Ciencias F\'{\i}sicas, UNAM}
\affiliation{Institute for Theoretical Physics, KU Leuven}
\author{Hern\'{a}n Larralde}
\email{hernan@fis.unam.mx}
\affiliation{Instituto de Ciencias F\'{\i}sicas, UNAM}
\author{Fran\c{c}ois Leyvraz}
\email{leyvraz@fis.unam.mx}
\affiliation{Instituto de Ciencias F\'{\i}sicas, UNAM}
\altaffiliation{Centro Internacional de Ciencias, Cuernavaca, Mexico}
\begin{abstract}

We consider the statistical mechanics of a small gaseous system
subject to a constant external field. As is well known, in the
canonical ensemble the system i) obeys a barometric formula for the
density profile and ii) the kinetic temperature is independent of
height, even when the system is small.  We show here that in the
microcanonical ensemble the kinetic temperature of the particles
affected by the field is not constant with height, but that rather,
generally speaking, it decreases with a gradient of order $1/N$. Even
more, if we have a mixture of two species, one which is influenced by
the field and the other which is not, we find that the two species'
kinetic temperatures are generally different, even at the same
height. These facts are shown in detail by studying a simple
mechanical model: a Lorentz Gas where particles and spinning disks
interact and the particles are subjected to a constant external force.
In the microcanonical ensemble, the kinetic temperature of the
particles is indeed found to vary with height; the disks' kinetic
temperature, on the other hand, is height-independent, and thus,
differs from that of the particles with which they interact.
\end{abstract}

\maketitle
\section{Introduction}
When considering small many-body systems one can expect
ensemble-dependence on the thermodynamic quantities \cite{LynBell,
  gross01,hill}. In most cases, it is not until the thermodynamic
limit is reached that the difference between statistical ensembles
disappears. For finite systems, however, differences between the
different statistical ensembles may be both significant and rather
intriguing. Although this has been known ever since the early stages
of statistical mechanics, the small-system-regime where ensemble
dependences matter has only become relevant to experiments and/or
applications during the last decades. Indeed, the microscopic and
mesoscopic scales are becoming more important due to technological
advances e.g., in nanosystems, as well as by the broadening scope of
physics toward phenomena from other disciplines of science, such as
molecular chemistry, micromechanics and biological systems. Thus, it
may be interesting to review and study the basic properties of simple
realistic models in this regime.

In this paper we study the equilibrium properties of a gas under the
influence of a constant external field. In the canonical ensemble the
system will display a {\em barometric}, that is, exponential,
dependence of the density on height; whereas the temperature, which in
the canonical ensemble can always be computed from the mean kinetic
energy of the particles, is independent of height. Our aim is to look
at the way in which these facts are modified when considering a
simple, yet realistic, isolated system.  The system we shall study is
the so-called Spinning Lorentz Gas (SLG) -- whose transport properties
have been presented in \cite{PRL24, SLG-JSP}. The SLG is simply the
Lorentz gas in which the circular scatterers are allowed to rotate and
exchange energy with the scattered particles; the system is described
in Section \ref{slgbarometric}.  This model has been shown to have
realistic transport properties: specifically, it displays normal
transport (Fourier's and Fick' laws hold) and it is well described by
the hypothesis of Local Thermal Equilibrium. It also shows coupled
mass and energy transport and satisfies Onsager's reprocity relations.
The SLG is thus a very simple interacting particle model with
reasonably realistic properties. Its merit in this paper is that its
equilibrium statistics can be solved exactly.

In what follows we study an isolated (microcanonical) SLG with a
constant external field acting on the particles.  We find in Section
\ref{micsys} that with this setup 
the particles reach a non uniform density profile resembling the
barometric formula (we note, parenthetically, that the usual Lorentz
gas model \cite{Lor05} does not reach such a barometric profile under
the effect of an external field).  We show there that the {\it kinetic
  temperature} of the particles varies with the height in the system,
while the kinetic temperature of the scatterers is constant. We show
as well that the kinetic temperature of the scatterers also differs
slightly from the proper temperature, calculated from the derivative
of the entropy with respect to energy. These effects disappear, of
course, in the thermodynamic limit.  These results highlight the fact
that when a finite and closed system is in equilibrium (in the
microcanonical ensemble), the question of how to identify the local
temperature must be answered carefully, as kinetic temperature
gradients can be present in the system in equilibrium.  We have also
performed molecular dynamics simulations of this system to verify our
results. We have found that indeed, the kinetic temperature of the
particles varies according to their height in the system, whereas that
of the scatterers is constant as predicted; however, since the applied
field affects the ergodicity properties of the system, microcanonical
Monte Carlo simulations were also performed. Next we argue in Section
\ref{gener} that such temperature gradients occur rather generally.
From a historical point of view, it may be of interest to note that
the height dependence of temperature in systems under the influence of
gravity was already stated by Loschmidt \cite{losch}. His results
strongly exaggerate the effect, however, and do not adequately
consider interactions.

\section{Barometric Lorentz Gas}
\label{slgbarometric}
\begin{figure}[hbt]
\begin{center}
\includegraphics{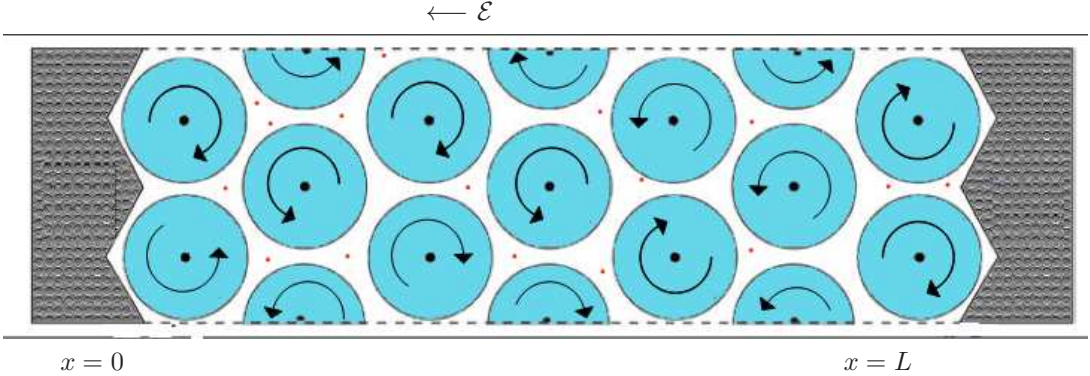}
\caption{The geometric setting of our SLG system: in the closed slab
  of length $L$ the array of $M$ rotors is set in a (finite horizon)
  triangular lattice so that $N$ particles (in dots) cannot enter and
  leave an hexagonal cell surrounding each disc without having at
  least one scattering collision. Particles are reflected elastically
  by walls at $x=0$ and $x=L$; the vertical coordinate is
  periodic. Two cells are the minimum width of the slab in order to
  avoid consecutive collisions with the same disk. The field strength
  $\elec$ is constant along the slab. }
\label{Pinball}
\end{center}
\end{figure}

In the barometric SLG, $N$ non-interacting particles of mass $m$ move
in a plane under the action of a constant applied force field of
strength $\elec$.  The particles can exchange energy with $M$ disk
scatterers which rotate freely, with their centers fixed in a (finite
horizon) triangular lattice. At the walls on either end of the system
(see figure \ref{Pinball}) particles are reflected elastically, while
the vertical coordinate $y$ has periodic boundary conditions. The
energy of the system is given by
\begin{equation}
H=\sum\limits_{i=1}^{N}\left(\frac{m}{2}\mathbf{v}_{i}^{2} +q\elec{}x_{i}\right) 
+\frac{\Theta }{2} \sum\limits_{i=1}^{M}\mathbf{\omega }_{j}^{2},
\label{Hslg}
\end{equation}
\\[-0.2cm] where $\mathbf{v}_{i}=\mathbf{p}_i/m$ is the velocity of a
particle with coordinate $x_i$, and $\omega _j$ is the angular
velocity of the $j^{th}$ scatterering disk, which have moment of
inertia $\Theta $.  Interactions in the SLG result from reversible,
energy conserving collisions between particles and disks. The
collision rules are given by
\begin{equation}
\begin{array}{c}
v_{n}^{\prime }=-v_{n},\qquad v_{\shortparallel }^{^{\prime
}}=v_{\shortparallel }-\frac{2\eta }{1+\eta }\left( v_{\shortparallel
}-R\omega \right) , \\ 
\\ 
R\omega ^{^{\prime }}=R\omega +\frac{2}{1+\eta }\left( v_{\shortparallel
}-R\omega \right) ,%
\end{array}%
\end{equation}%
where $v_{n,\shortparallel }$ are normal and tangential components of
the particle's velocity with respect to disk surface, $R$ is the disk
radius and $\eta \equiv \Theta /mR^{2}$ is the dimensionless parameter
that controls the fraction of total energy exchanged between the disk
and the particle \cite{PRL24}.

Finally, note that this system has {\em two\/} different confining
mechanisms: on the one hand the field limits effectively the particles
in a finite region, on the other, there exists a finite box of size
$L$. In general the case in which $L\to\infty$ is significantly
easier, and we shall ocasionally sketch the derivations in that case.

\section{Definition of Temperature}

Classically, in the microcanonical ensemble, up to an additive
constant one has the following expression for the entropy $S(E)$ as a
function of the energy $E$
\begin{eqnarray}
    S(E)&=&k_{B}\ln \Omega(E)\nonumber\\
    &=&k_{B}\ln\int_{\Gamma}d^{N}\vec p\,d^{N}\vec 
    q\, \delta\left[E-H(\vec p, \vec q)\right]
    \label{eq:new1}
\end{eqnarray}
where $\Gamma$ is the phase space of the system and $H$ is its
Hamiltonian. The temperature is then given by
\begin{equation}
    \frac{1}{k_{B}T_M}=\frac{\partial S(E)}{\partial E}
    \label{eq:new2}
\end{equation}
Here $T_M$ stands for the microcanonical temperature as defined by (\ref{eq:new2}).
If $H$ is of the form $\sum p_i^2/2m_i+ V(q_1,...)$, as it is in our
case, it is readily shown that \cite{Pears}
\begin{equation}
    \frac{1}{k_{B}T_M}=\left(
    \frac{dN}{2}-1
    \right)\left\langle
    K^{-1}
   \right\rangle
    \label{eq:new2.5}
\end{equation}
where $K$ stands for the total kinetic energy of the system, $d$ is
the dimensionality of ambient space and $N$ is the number of
particles.  In contrast, in the canonical ensemble, one has the
relation
\begin{equation}
    k_{B}T_C=\frac{2}{dN}\left\langle
    K
    \right\rangle,
    \label{eq:new3}
\end{equation}
which we will refer to as the {\it kinetic temperature}.  If the
distribution of $K$ is strongly peaked at one value, as usually
happens for large systems, both definitions \refeq{eq:new2.5} and \refeq{eq:new3} 
will coincide.  This is a
special instance of ensemble equivalence in the thermodynamic limit,
and is to be expected on general grounds except for special cases
\cite{review,abelardo,abelardo1}.  On the other hand, the two definitions of temperature will
generally differ in finite systems.  Nevertheless, since the system we
consider can be thought of as being made up of subsystems in contact
with each other, it seems reasonable that the ``local temperature''
could be identified with the average kinetic energy of particles at
each position. In a sense, the main message of this paper is precisely that such 
intuition is altogether untenable, at least in the microcanonical ensemble
if one considers terms of order $1/N$. 

\section{Microcanonical calculation}
\label{micsys}

The statistics in the microcanonical ensemble are given by 
\begin{equation}
\rho_E(\vec p_i, \omega_i;x_i, y_i,
\phi_i)):=\frac{1}{\Omega_{N,M}(E)}\delta\left[ E-H(\vec p_i,
  \omega_i;x_i, y_i, \phi_i)\right]
\label{eq:2}
\end{equation}
where $\Omega_{M,N}(E)$ is the microcanonical partition function,
with the dependencies on $E$, $M$ and $N$ explicitly displayed 
\begin{equation}
\Omega_{N,M}(E)=\int
d^{2N}p\,d^Nx\,d^Ny\,d^M\omega\,\delta\left[E-H(\vec p_i,
  \omega_j;x_i, y_i, \phi_j)\right]
\label{omega}
\end{equation}
and the Hamiltonian we consider is given by
\begin{equation}
H(\vec p_i, \omega_j;x_i, y_i, \phi_j)=\sum_{i=1}^N\frac{p_i^2}{2m}
+\frac1{2\Theta}\sum_{i=1}^M\omega_i^2+q\elec{}\sum_{i=1}^N x_i
\label{eq:1}
\end{equation}
What we want to compute is the kinetic temperature at $x_0$.
That is, we want to know the average kinetic energy of a particle,
given that its position is in the infinitesimal interval
$[x_0, x_0+ d x_0]$.  The local kinetic temperature is expressed as
\begin{equation}
k_B T(x_0)=\int dp_x\,dp_y\,\frac{p_x^2+p_y^2}{2m}\rho(\vec p|x_0),
\label{eq:3}
\end{equation}
where $\rho(\vec p|x_0)$ is the momentum distribution conditional to
the particle position being $x_0$. This conditional probability is, of
course, normalized, i.e.
\begin{equation}
\int d^2\vec p_0\,\rho(\vec p_0|x_0)=1,
\label{eq:4}
\end{equation}
it follows, then, that if $\rho(p_0|x_0)$ is known up to an arbitrary
multiplicative constant which is independent of $p_0$, then it is in
fact fully known since (\ref{eq:4}) determines this constant. Thus,
we may without loss of information discard any multiplicative constant
that is independent of $p_0$. Denoting by $\propto$ the equality of two
expressions up to such a constant we have
\begin{eqnarray}
\rho(\vec p_0|x_0)&\propto&\int
d^{2N}p\,d^Nx\,d^Ny\,d^M\omega\,\rho_E(\vec p_i, \omega_i; x_i, y_i,
\phi_i))\,\times\nonumber\\ &&\quad\times\sum_{j=1}^N\delta(\vec
p_j-\vec p_0)\,\delta(x_j-x_0).
\label{eq:7}
\end{eqnarray}
Thus
\begin{equation}
\rho(\vec p_0|x_0)\propto\Omega_{N-1,M}\left(E-q\elec{}x_0-\frac{p_0^2}{2m}
\right)
\label{eq:7.5}
\end{equation}
To proceed we must calculate $\Omega_{N,M}(E)$.  In the case $L\to\infty$, 
an entirely straightforward scaling argument shows that 
\begin{equation}
\Omega_{N,M}(E)\propto E^{2N+M/2-1}
\label{eq:7.5.1}
\end{equation}
From which the kinetic temperature in this limit, (\ref{TLargesyst}),
can be readily derived.  We proceed directly to the general case, which is
slightly more involved: consider the Laplace transform
\begin{equation}
{\hat \Omega}_{N,M}(z)=\int\limits_0^\infty e^{-zE}\Omega_{N,M}(E) dE
\end{equation}
or, explicitly:
\begin{equation}
{\hat \Omega}_{N,M}(z)=\int\limits_{-\infty}^\infty
d^{2N}p\,\int\limits_{0}^L d^Nx\,\int\limits_{0}^W
d^Ny\,\int\limits_{-\infty}^\infty
d^M\omega\,\exp\left[
-z
\left(
\sum_{i=1}^N\frac{p_i^2}{2m}
+\frac1{2\Theta}\sum_{i=1}^M\omega_i^2+q\elec{}\sum_{i=1}^N x_i
\right)
\right]
\end{equation}
where $L$ is the length of the system in the $x-$direction and $W$ is
the length in the periodic $y-$direction. The advantage of this
expression is, of course, that the integrals are separable and can be
evaluated. Again omitting irrelevant constants, we have:
\begin{equation}
{\hat \Omega}_{N,M}(z)\propto z^{-2N-M/2} \left(1-e^{-zq\elec{}L}\right)^N.
\label{eq:omega-laplace}
\end{equation}
In order to state the final results more expeditiously, we first define a function $\Phi_N(\nu|x)$
as follows:
\begin{equation}
\Phi_N(\nu|x)=\sum_{n=0}^{N}\left(
\begin{array}{c}
N\\
n
\end{array}
\right)
(-1)^n(x-n)^\nu\heavi(x-n),
\label{eq:defPhi}
\end{equation}
where $\heavi(x)$ is the step function. 

It is now readily seen that, by inverting (\ref{eq:omega-laplace}), we obtain
\begin{equation}
{\Omega}_{N,M}(E)\propto
(q\elec{}L)^{2N+M/2-1} \Phi_N
\left(
2N+M/2-1\left|\frac{E}{q\elec{}L}
\right)\right.
.
\label{omega-new}
\end{equation}
We can now calculate
$\rho(\vec p_0|x_0)$ explicitly:
%
\begin{equation}
\rho(\vec p_0|x_0)=\left(
\frac{2N+M/2-2}{2\pi mq\elec{}L}
\right)
\frac{\Phi_{N-1}\left(
2N+M/2-3\left|
\frac{E-q\elec{}x_0-p_0^2/(2m)}{q\elec{}L}
\right)
\right.}{\Phi_{N-1}\left(
2N+M/2-2\left|
\frac{E-q\elec{}x_0}{q\elec{}L}
\right)\right.}
\label{eq:explicit}
\end{equation}

Thus, the kinetic temperature as a function of position $x_0$ is given by
%
\begin{equation}
k_B T(x_0)=\left(
\frac{q\elec{}L}{2N+M/2-1}
\right)
\frac{\Phi_{N-1}
\left(
2N+M/2-1
\left|
\frac{E-q\elec{}x_0}{q\elec{}L}
\right)\right.
}{\Phi_{N-1}
\left(
2N+M/2-2\left|
\frac{E-q\elec{}x_0}{q\elec{}L}
 \right)
\right.}
\label{eq:Tprofile}
\end{equation}

We observe in \refeq{eq:Tprofile} that for finite values of $N$ and $M$, if the
temperature $T(x)$ is defined as the local average of the kinetic
energy of the particles, then such temperature is {\em not constant\/}
as a function of $x$. The limiting behaviors of the above expression
are relatively easy to evaluate. First we consider the case in which
the system size grows to infinity: when $q\elec{}L>E$ then the step
functions are zero for all value of $x_0$, except for the term
$n=0$. Thus, in this limit we have
\begin{eqnarray}
k_B T(x_0)=\frac{(E-q\elec{}x_0)~\heavi(E-q\elec{}x_0)}{2N+M/2-1}.
\label{TLargesyst}
\end{eqnarray}
It is amusing to note that this may be an outrageous limit: if the
field is earth's gravity, for example, to achieve the desired limit
we need heights larger than those that would be reached if we allocate
{\it all} the energy of the system as the gravitational potential
energy of a single particle. If we consider that molecular masses are
of the order of $10^{-25}$ kg, energies corresponding to temperatures
of a few degrees Kelvin correspond to heights of hundreds of meters
even for systems consisting of a few molecules.

The limit in which the field vanishes $Nq\elec{}L/E \to 0$, can be calculated 
using the fact that
\begin{equation}
\sum\limits_{m=0}^M \frac{M!}{m!(M-m)!} (-1)^m m^n=
(-1)^M\begin{cases} 0 & {\rm if}\qquad n<M; \\ M! & {\rm if} \qquad
n=M; \\ M(M+1)!/2 & {\rm if} \qquad n=M+1; \\ \qquad\vdots\\
\end{cases}
\end{equation}
Then, in this limit, the kinetic temperature of the particles becomes
\begin{eqnarray}
k_B T(x_0)=\frac{E}{N+M/2}\left[1 - \frac{q\elec{}NL}{E}\left(1-\frac{(L-x_0)}{NL}\right)+\cdots\right];
\qquad (Nq\elec{}L/E \to 0)
\label{TLargeE}
\end{eqnarray}
where we have kept terms to leading order in the strength of the field
only to highlight the first order at which the dependence on the
position $x_0$ appears.

We can make an entirely similar computation for the disks, where now
we want to calculate $g(w_{j})$, the probability that the $j^{th}$
scatterer has angular velocity $w_{j}$,
\begin{eqnarray}
    g(w_{j})\propto \Omega_{N,M-1}\left(
     E-\frac{\Theta w_{j}^2}{2}
     \right)
    \label{eq:new6}
\end{eqnarray}
Using the explicit expression for $\Omega_{N,M}(E)$ and normalizing, one obtains 
\begin{equation}
g(w_j)=2\left(\frac{\Theta}{2}\right)^{1/2}
\frac{\Gamma(2N+M/2)(q\elec{}L)^{-1/2}}{\Gamma(1/2)\Gamma(2N+M/2-1/2)}
\frac{\Phi_N
\left(
2N+M/2-3/2
\left|
\frac{E-\Theta w_j^2/2}{q\elec{}L}
\right)\right.}
{\Phi_N
\left(
2N+M/2-1
\left|
\frac{E}{q\elec{}L}
\right)
\right.}.
\label{explicit2}
\end{equation}

Thus, for the scatterers, using the mean kinetic energy to define the
temperature yields
\begin{equation}
k_BT_S(E)=\frac{q\elec{}L}{2N+M/2}\,
\frac{\Phi_N
\left(
2N+M/2
\left|
\frac{E}{q\elec{}L}
\right)\right.}
{\Phi_N
\left(
2N+M/2-1
\left|
\frac{E}{q\elec{}L}
\right)
\right.}
\label{explicit2.1}
\end{equation}
which, in contrast to \refeq{eq:Tprofile} is constant throughout the system.  In the limit $L\to\infty$,
again only the $n=0$ terms contribute and the above expression
becomes:
\begin{eqnarray}
k_B T_S(E)=\frac{E}{2N+M/2};
\label{TslargeL}
\end{eqnarray}
whereas in the limit $Nq\elec{}L/E \to 0$ one recovers the value
\begin{eqnarray}
k_BT_S(E)=\frac{E}{N+M/2}
\qquad (Nq\elec{}L/E \to 0).
\label{TslargeE}
\end{eqnarray}
Of course, in this limit the field $\elec$ no longer plays a role and the kinetic 
temperature of disks \refeq{TslargeE} and particles \refeq{TLargeE} tend to the same value.

Finally, to finish muddling the situation, we can calculate the 
temperature directly from eq. \refeq{eq:new2}, 
\begin{equation}
k_BT_M(E)=\frac{q\elec{}L}{2N+M/2-1}\,
\frac{\Phi_N
\left(
2N+M/2-1
\left|
\frac{E}{q\elec{}L}
\right)\right.}
{\Phi_N
\left(
2N+M/2-2
\left|
\frac{E}{q\elec{}L}
\right)
\right.}
\label{Tmexplicit}
\end{equation}
which is a constant of the system, albeit, not equal to the
kinetic temperature of the scatterers \refeq{explicit2.1}. Expression \refeq{Tmexplicit} takes the value
\begin{eqnarray}
k_B T_M(E)=\frac{E}{2N+M/2-1},
\end{eqnarray}
as $L\to\infty$, which again differs slightly from the value reached by the
scatterers \refeq{TslargeL}.  This is slightly unexpected: indeed, we might have expected that,
since the scatterers are in contact with the particles, these could be assimilated to
a ``thermal bath'', so that the scatterers would effectively be in the canonical ensemble. 
This is in fact correct if $N\gg M$ or if $M\gg1$, but not in general. 

On the other hand, in the limit, $Nq\elec{}L/E \to 0$, from \refeq{Tmexplicit} we obtain
\begin{eqnarray}
k_BT_M(E)=\frac{E}{N+M/2-1},
\qquad (Nq\elec{}L/E \to 0).
\end{eqnarray}
This difference is easy to understand, if we realize that 
the height now drops out as a variable which contributes to equipartition.

Of particular interest is the fact that the disks have a constant
kinetic temperature that is {\em different} from the local kinetic
temperature of the particles with which they interact. Further, the
global temperature in the system does not coincide with the kinetic
temperatures of the components. While these results appear to be quite
contrary to the usual notion of equilibrium, they are, in fact, a
consequence of the equilibrium statistics in the microcanonical
ensemble for finite systems.

In hindsight, the origin of the variation of the kinetic energy with
the height of the particles in system is easy to understand: The
presence of the applied field implies that potential energy is
required for particles to reach certain height. Since the total energy
is fixed, there is less energy left over to distribute amongst the
rest of the elements in the system. This is not the case for the
rotators, since there is no energy cost for their location in the
channel. Thus the kinetic temperature of the particles decreases with
height whereas that of the rotators remains constant. Still, it is
amusing to note that in spite of having different kinetic
temperatures, or precisely because they have different kinetic
temperatures, the scatterers and the particles are in equilibrium with
each other.  We present a sketch of the kinetic effects involved in
this apparent failure of collisions to yield equipartition of kinetic
energy in Appendix \ref{appkinet}.

For completeness, we also calculate the particle density, which can be
expressed as:
\begin{equation}
C(x_0)\propto \int\limits_{-\infty}^\infty d^2 p_0 \,
\Omega_{N-1,M}\left(E-q\elec{}x_0-\frac{p_0^2}{2m}
\right).
\label{density1}
\end{equation}
$C(x_0)$ must now be normalized to $N$, the number of particles
in the system. Using Eq.\ref{omega}, we obtain:
\begin{equation}
C(x_0)=\frac{N(q\elec{}L)^{2N+M/2-2}}{{\mathcal N}}\,
\Phi_{N-1}
\left(
2N+M/2-2
\left|
\frac{E-q\elec{}x_0}{q\elec{}L}
\right)\right.
\end{equation}
where the normalization constant ${\mathcal N}$ is
\begin{subequations}
\begin{eqnarray}
{\mathcal N}&=&\frac{1}{q\elec{}(2N+M/2-1)}
\left(
E^{2N+M/2-1}+S_1+S_2
\right)\\
S_1&=&\sum\limits_{n=1}^{n*}
\frac{(N-2n)(N-1)!}{n!(N-n)!}(E-nq\elec{}L)^{2N+M/2-1}\\
S_2&=&\sum\limits_{n=n*}^{N-1}
\frac{(N-1)!}{n!(N-n-1)!}(E-nq\elec{}L)^{2N+M/2-1}
\end{eqnarray}
\end{subequations}
where $n*=Int[\frac{E}{q\elec{}L}]$. The convoluted expression for
${\mathcal N}$ arises from the fact that the integral of some terms of
the sum are cut by the step function in the expressions, whereas
others are cut by the finite size $L$ of the system.  It is, of
course $C(x_0)$,  the density profile of the particles, which
becomes the familiar exponential in the thermodynamic limit.

\section{General Systems}
\label{gener}

While the explicit calculations presented above apply directly to the SLG, we
argue that the effects illustrated with this model are rather
general. Our observations rest essentially on (\ref{eq:7.5}) and
(\ref{eq:new6}), in which we express the probability of finding a
particle, respectively a disk, with a given kinetic energy in terms of
the microcanonical partition function. The reasoning leading to these
equations is, of course, entirely general.

If we take an arbitrary system, it is necessary to resort to
approximations, but the result is quite similar to the ones obtained
for the SLG. For the sake of simplicity, we limit ourselves here to a
finite number $N$ of particles {\em only\/} confined by the field,
that is, we neglect the effect of a confining box altogether. One has
in the general case, for the particles subjected to the field:
\begin{eqnarray}
\ln\rho(\vec p_0|x_0)&=&
\ln\Omega_{N-1,M}\left(E-q\elec{}x_0-\frac{p_0^2}{2m}\right)-\ln\Omega_{N-1,M}(E)+K
\nonumber\\
&\approx&K-
\left(
q\elec{}x_0+\frac{p_0^2}{2m}
\right)\frac{\partial}{\partial E}
\ln\Omega_{N,M}(E)
+\frac12\left(
q\elec{}x_0+\frac{p_0^2}{2m}
\right)^2
\frac{\partial^2}{\partial E^2}
\ln\Omega_{N,M}(E)
\nonumber\\
&=&K^\prime-\frac{p_0^2}{2mk_BT_M}
-
\frac1{2C_Vk_BT_M^2}\left(
q\elec{}x_0+\frac{p_0^2}{2m}
\right)^2\nonumber\\
&=&K^{\prime\prime}-\frac{p_0^2}{2mk_BT_M}
\left(
1
+
\frac{q\elec{}x_0}{C_VT_M}
\right)-
\frac{p_0^4}{8m^2C_Vk_BT_M^2}
\label{eq:7.5new}
\end{eqnarray}
where $K$ and its primed variants denote additive constants
independent of $p_0$, corresponding to the undetermined multiplicative
constant in (\ref{eq:7.5}), and $C_V$ is the heat capacity at
constant volume. The second term clearly shows that the temperature
has the kind of dependence stated in this paper. Indeed
\begin{equation}
\frac{dT(x_0)}{dx_0}=-
\frac{q\elec{}}{C_V}
\label{Tprofilegen}
\end{equation}
up to terms of higher order in $1/N$.  The third term in
(\ref{eq:7.5new}), on the other hand, indicates a deviation from the
Maxwellian in microcanonical systems at the $1/N$ level. It thus
generates a correction to $T$ of order $1/N$ but independent of $x_0$.

If we additionally have another species which is not affected by the
external field, its kinetic temperature will be unaffected by the
field and thus independent of $x$. This kinetic temperature of this
species will differ in general from the kinetic temperature of the
other species. We thus see that quite generally the kinetic
temperature neither equilibrates between different heights, nor
between different species.  On the other hand, the microcanonical
temperature is a characteristic of the whole system, but, contrary to
the kinetic temperature, there is no clear way of attributing it to
any part of the system, such as a species, or a position.

\section{Numerical Simulations}

To check the validity of our results, we have performed extensive
molecular dynamics simulations of the system as well as Monte Carlo
simulations, since with the latter no problems arise with the sampling
of the $N$-particle phase space.  For all the simulations, particle
and disk masses were set to one, $m=M=1$, as well as the interaction
parameter $\eta=1$, which controls the energy exchange between
particles and scatterers. To calculate local averages, the channel of
Fig. \ref{Pinball} is divided in $\boldsymbol{\mathcal{S} }$
``strips'' of width $\Delta x=L/\boldsymbol{\mathcal{S}}$, where $L$
is the total channel length.

The density $C(x)$, and kinetic temperature $T_K(x)$ were obtained
as the time average of particle number per area and the
time average of the kinetic energy per particle in each strip.

\begin{figure}[htb]
\includegraphics{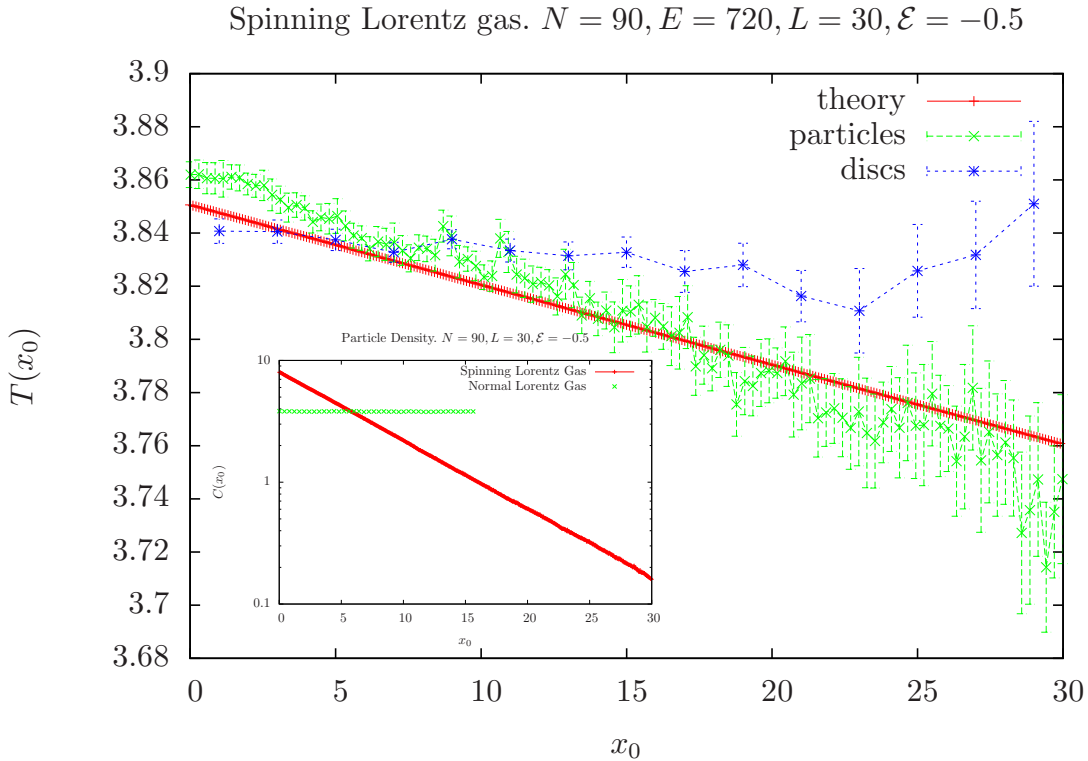}
\caption{Kinetic temperature profile in the SLG for $N=90$ particles
  inside the closed slab with length $L=30$ (see Fig.~\ref{Pinball}),
  with an external applied field $\elec=-0.5$ and system energy of
  $E=720$. These results were obtained by Molecular Dynamics
  simulation (segmented lines). The continuous line (red) is the
  theoretical prediction (\ref{eq:Tprofile}). We observe a slight
  discrepancy for $T_K(x)$ along the slab; the reason is presumably
  related to issues of non-ergodicity of the simulation. The
  short-segmented line (stars) indicates the temperature of the
  discs. These data are also significantly more noisy than in the
  Monte-Carlo simulations. In the inset, we show a semi-log plot of
  the density of particles $C(x)$ in the slab, compared to a similar
  simulation in the usual Lorentz gas (in crosses).}  
\label{kineptMD}
\end{figure}

\begin{figure}[htb]
\includegraphics{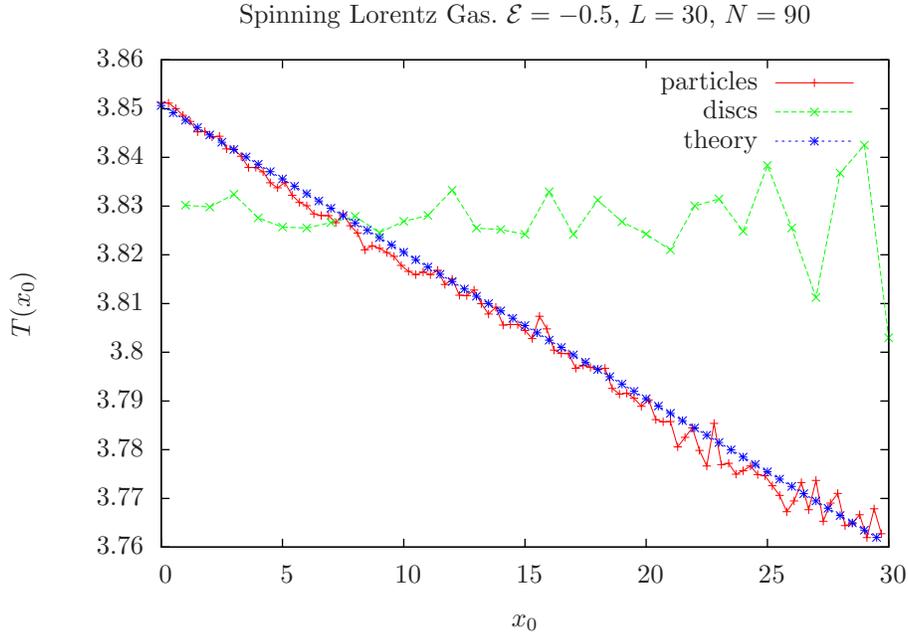}
\caption{This figure shows results form a Monte-Carlo simulation of
  the SLG model to obtain the kinetic temperature profile for $N=90$
  particles inside a closed slab of length $L=30$ with an external
  applied field $\elec=-0.5$ and an initial energy of $E=720$. The
  continuous line shows the simulation data while the discontinuous
  line (stars) is the theoretical prediction
  (\ref{eq:Tprofile}). There are thus no approximations involved.  The
  discontinuous line with crosses (green) indicates the temperature of
  the discs.  }
\label{kineptMC}
\end{figure}

In Fig.~\ref{kineptMD} we show that $T_K(x)$ is indeed not constant
for our closed, many-particle system, as expected from our
results. However, the kinetic temperature measured in the molecular
dynamics simulations appears to display a very slight, but possibly
systematic, deviation from the value obtained analytically. We believe
that such discrepancies may arise from the fact that the presence of
the field affects the ergodicity properties of the system. In
particular, some configurations are hard to reach from generic initial
conditions; for example in those for which the disks have a large
share of the energy, particles become confined to the region near the
bottom of the channel. To check whether this was the case, we
performed microcanonical Monte Carlo simulations, in which particles
were allowed to fly under the influence of the field, and after a
certain amount of time, they would either exchange energy with a
random scatterer or randomly rotate their velocity vector. The kinetic
temperatures for this simulation agree quite well with the theoretical
prediction (see Fig.~\ref{kineptMC}).

The density of particles $C(x)$ is also shown in Fig.~\ref{kineptMD}
(inset); it was measured for both the SLG and the normal Lorentz gas
for similar simulations with an applied field $\elec =-0.5$, where all
particles are initially at height $x=15$ while discs and particles
have zero kinetic energy. We observe that in the SLG the density
profile is indeed {\it barometric}; in contrast, in the normal Lorentz
gas particles cannot go beyond their maximum initial energy: it is
therefore impossible to obtain such a barometric profile without some
kind of interaction, as provided for example by the SLG model.

\section{Conclusion}

Summarising: in isolated systems described by the microcanonical
ensemble, the presence of an applied field gives rise to intrinsic
equilibrium inhomogeneities: a spatially varying local kinetic
temperature which differs from the thermodynamic temperature of the
system. We argue that this result is general, and we illustrate the
effect both analytically and numerically for the SLG model, subject to
a constant external field, for which all calculations can be carried
out explicitly.  The effect vanishes in the limit $N\to\infty$, but
only as $1/N$, so that it may be observable in small systems.

If we have another species in the system which is not affected by the
external field, its  kinetic temperature will be unaffected by
the field. The kinetic temperature of this species will be quite close
to the microcanonical temperature of the whole system, at least if the
particle numbers are not too small. On the other hand, the two 
species' kinetic temperatures will not equilibrate, so that we cannot
identify it as the (local) {\em thermodynamic\/} temperature of the species.

Further, an interesting possibility should be pointed out: when small
systems are considered in the microcanonical ensemble, the possibility
of negative specific heat cannot be ruled out \cite{review, abelardo,
abelardo1}, particularly if the system has long range interactions
and is close to a tricritical point. We therefore cannot exclude the
possibility that a temperature gradient arises in which high lying
particles actually have {\em higher\/} temperatures than low lying
ones.

 These results must be all carefully considered 
in any applications of statistical thermodynamics 
to small many-particle systems such as e.g. atomic clusters, nanoparticles 
or molecular/biological ensembles, 
since these systems may not necessarily 
be described using the thermodynamic limit.

\begin{acknowledgements}
One of us (FL) would like to thank UNAM PAPIIT grant IN 114014
as well as CONACyT grant 154586 for financial support. HL acknowledges
the support of CONACyT grant 129471. 
\end{acknowledgements}

\appendix

\section{Kinetic observations on the failure of disks to equilibrate with particles}
\label{appkinet}
The conundrum arising from the behaviour of the scatterers may be best
understood in the light of the following remarks: First, a high lying
scatterer can have a kinetic temperature which is significantly higher
than that of the particles in the system. Second: most of the time
such a scatterer has no particle in its vicinity. Finally, if we
calculate the mean kinetic energy of the scatterer {\em conditional\/}
on the presence of a particle in the same cell, it is the same as the
particle temperature at this height.  The presence of several
particles, as an entirely similar calculation to the ones performed
here shows that the temperature of the scatterer will be still lower.

This leads to the conclusion that the scatterer must be significantly
hotter when it finds itself in the absence of any particle than in the
presence of one or more particles.  Since the mere absence of
particles cannot of itself heat up the scatterer, we must find a
mechanism whereby the scatterer is preferentially heated to an
anomalous extent at the precise moment when the last particle leaves
the scatterer's vicinity.

To this end, let us use an exceedingly simplified model of what takes
place in the system. Consider two cells, an upper and a lower
one. Each cell contains a scatterer, which always remains in
the cell and only has kinetic energy.
 
Additionally the whole system contains one particle, which can
alternate between the two cells, and which has both kinetic and
potential energy, the latter being always $V/2$ in the upper cell and
$-V/2$ in the lower.  Such a system in the microcanonical ensemble has
all the features we look for. In particular, the particle in the upper
cell is significantly colder than the corresponding scatterer. Indeed,
in that case, all the components have the same energy, which leads to the
kinetic energy of the particle being more than that of the scatterer
in the lower and less in the upper cell.

In order to generate the microcanonical ensemble, we use the following
Monte-Carlo dynamics: at each step, with probability $\epsilon$ a move
of the particle from one cell to the other is attempted. If the
particle is in the upper cell, the move is always accepted and the
particle's kinetic energy is increased by $V$. Otherwise, the move is
only accepted if the total energy of the particle is larger than $V$,
The kinetic energy is then decreased by $V$. With probability
$1-\epsilon$, on the other hand, the particle and the scatterer in the
particle's cell add up their kinetic energies and then proceed to
redivide the total energy randomly. As is readily seen, this Markov
process satisfies detailed balance with respect to the microcanonical
ensemble and thus tends to it, at least if ergodicity is satisfied.

This systems displays exactly the kind of ``paradox'' described in
this paper: on average, the scatterer has a different kinetic energy
than the particle, yet, when $\epsilon$ is small they pass a long time
together and equilibrate their kinetic energies, thus it is not clear
intuitively from what effect the discrepancy in temperatures could
arise. Indeed, it is not difficult to show in this simplified model,
that the kinetic energy of a scatterer {\em conditional\/} on the
presence of the particle in the cell, is indeed close to the particle
kinetic energy.

Thus the only way in which the discrepancy can arise is in the last
exchange of energy just before the particle changes cell. Indeed, if
the particle leaves the upper cell for the lower one with an
exceptionally small amount of kinetic energy, then  the
scatterer in the upper cell keeps an anomalously large amount of
kinetic energy. Further, the particle in these circumstances will take
a longer than normal time to come back to the upper cell. Thus the
upper scatterer will have remained for an anomalously long time in a
state of anomalously high kinetic energy. The repetition of this
pattern is, as can be checked, sufficient to cause a finite difference
in the kinetic energies of the upper scatterer and the particle in the
upper cell.

\end{document}